\documentstyle[sprocl,epsf]{article}

\def\Journal#1#2#3#4{{#1} {\bf #2}, #3 (#4)}
 
\def\NIMA{{\em Nucl. Instrum. Methods} A}

\def\PRD{{\em Phys. Rev.} D}

\begin{document}

\title{SPIN EFFECTS IN RADIATING COMPACT BINARIES}
 
\author{ L\'{A}SZL\'{O} \'{A}. GERGELY$^{1,2}$, ZOLT\'{A}N
PERJ\'{E}S$^{1}$ and M\'ATY\'AS VAS\'UTH$^{1}$ }
 
\address{$^{1}$KFKI Research Institute for Particle and Nuclear
Physics, \\Budapest 114, P.O.B 49, H-1525 Hungary\\
$^{2}$Laboratoire de Physique Th\'{e}orique, Universit\'{e} Louis Pasteur,\\
3-5 rue de l'Universit\'{e} 67084 Strasbourg Cedex, France}
 
\maketitle\abstracts{
We review and summarize our results concerning the influence of the spins of
a compact binary system on the motion of the binary and on its gravitational
reaction. We describe briefly our method which lead us to compute the
secular changes in the post-Newtonian motion and the averaged radiative
losses. Our description is valid to 1.5 post-Newtonian order. All
spin-orbit and some spin-spin effects are considered which contribute
at this accuracy. This approach enabled us to give both the evolutions of the
constants of the nonradiative motion and of the relevant angular variables
under radiation reaction.
}

\section{Introduction}

Gravitational radiation, predicted by Einstein's theory, has long been
unavailable to experimentalists because of the low power of laboratory
sources. However, a new generation of earth-based gravitational-wave
detectors is approaching its final stage of construction (LIGO
\cite{LIGO}, VIRGO \cite{VIRGO}, GEO600 and TAMA300). 
A strong hope is that in the next decade direct experimental evidence
for this brilliant theoretical prediction will be obtained.
 
Coalescing binary neutron star systems are certainly among the
most promising sources for earth-based detectors with
the frequency range ($1-10^4$ Hz). Neutron star-black
hole and black hole-black hole binaries are also significant
sources \cite{Thorne} for the frequency range ($10^{-4}-1$ Hz)
of the Laser Interferometer Space Antenna (LISA). For data
reduction, signal
templates as well as the knowledge of the reaction effects of the
gravitational radiation emitted by these compact binary systems
are needed to a high precision.
There are indications that computations up to the
third post-Newtonian (3PN) approximation will ensure the required
accuracy. The computations have almost reached this level;
there are several generic treatments at 2PN accuracy \cite{Blanchet,WW,GI}
and a recent one \cite{BFP} at 2.5PN, 
however in most cases spin effects were not taken into account.
Binary systems do, in many cases, have a
non-negligible spin. In a series of papers \cite{GPV1,GPV2,GPV3} 
we have considered the influence of the spin on
radiation reaction.
 
Spin-orbit and spin-spin effects appear at 1.5PN and 2PN orders,
respectively. The instantaneous losses in the constants
characterizing the nonradiative motion (the energy and the
total angular momentum vector) and also the wave forms in the
presence of these spin effects were given by Kidder \cite{Kidder}
for generic eccentric orbits.
He has used the Blanchet-Damour-Iyer formalism\cite{BDI}
for evaluating the symmetric trace-free moments,
the covariant spin supplementary condition ({\em SSC})
and employed a description of the binary motion following Barker
and O'Connell \cite{BOC}, Thorne and Hartle \cite{TH} and
Kidder, Will and Wiseman \cite{KWW}. He gave also the
averaged losses of the dynamical quantities for circular orbits.
 
Despite the classical result on the circularization of the orbits
due to gravitational radiation reaction, eccentric orbits can
be relevant in various physical scenarios, as emphasized by several
authors \cite{GI,ShaTe,QuSha,HiBe}. Such binaries are likely to be
formed, for example, in galactic nuclei by capture events, in which
time is insufficient for circularization before plunging.
 
Averaging over eccentric orbits however turns out to be difficult
for binary systems with spins. We are content to include
leading order spin effects, which appear at 1.5PN order.
For a test particle,
this is a good approximation either for a black hole-neutron
star binary or for the debris particle orbiting about a massive
spinning central body. The averaged losses in the constants
characterizing the nonradiative motion on eccentric orbits were
given by Ryan\cite{Ry2}. His analysis lead to the same results as
our approach based on the Lense-Thirring picture \cite{GPV1}.
 
\section{The method and the results}
 
We have generalized our description for the case when the masses
of the two bodies are comparable, but one spin dominates over
the other \cite{GPV2} and, more recently, for comparable
masses and spins \cite{GPV3}.
 
For a full description to the 1.5PN order we have
introduced additional angle variables (Fig.1), which are not constant
even in the absence of radiation, and computed their radiative changes.
These angles subtended by the directions of the Newtonian orbital
angular momentum ${\bf{\hat L}_N}$ and spin vectors ${\bf{\hat S}_i}$
can be important in monitoring the relative orientation
of the binary with respect to the detector. For circular orbits,
the evolution of these angles has been discussed in recent
works \cite{ACST,Apost,deFelice}. For eccentric
orbits we have given both the instantaneous and averaged
evolution equations. We have found that eccentricity speeds up the
evolution.
 
In order to carry out these computations we had to appeal to the
Burke-Thorne \cite{BT} potential, since in the angular losses, the
radiative losses of the spins give contributions.
 
A striking feature of these losses was that (although these
Burke-Thorne potential terms are present in the instantaneous
losses) they average out to
\twocolumn\noindent
zero in the corresponding secular
expressions. Our results concerning
the losses in the constants of motion are in agreement with results
of Rieth and Sch\"afer \cite{RS}, which were obtained in
an other SSC. When complemented with our equations for the
angles in terms of the semimajor axis $a$, mass ratio $\eta=m_2/m_1$
and eccentricity $e$,
\begin{eqnarray}
\left\langle \!\frac{d\kappa_1}{dt}\!\right\rangle\!\! =\!\!
\frac {G^{7/2}m^{3/2}\mu}{30c^7a^{11/2}(1\!-\!e^2)^4} \Bigl\{
 (285e^4\!\!+\!1512e^2\!\!+\!488)
   (S_1\sin\kappa_1\!+\!S_2\sin\kappa_2\cos\Delta\psi) \nonumber\\
\!+\! (221e^4\!+\!1190e^2\!+\!384)
 (\eta S_1\sin\kappa_1\!+\!\eta^{\!-\!1}S_2\sin\kappa_2\cos\Delta\psi)
                                                 \nonumber\\
\!+\! (156e^4\!+\!240e^2)
  (S_1\sin\kappa_1\cos(2\psi_1\!-\!2\psi_0)
    \!+\!S_2\sin\kappa_2\cos(\psi_1\!+\!\psi_2\!-\!2\psi_0)) \nonumber\\
\!+\! (119e^4\!+\!193e^2)
 (\eta S_1\sin\kappa_1\cos(2\psi_1\!-\!2\psi_0)
   \!+\!\eta^{\!-\!1}S_2\sin\kappa_2\cos(\psi_1\!+\!\psi_2\!-\!2\psi_0)) \nonumber
\end{eqnarray}
we have a complete dynamical system describing the
evolution of the radiating binary \cite{GPV3}.
\begin{figure}[htb]
\epsfysize=6cm
\centerline{\hfill
\epsfbox{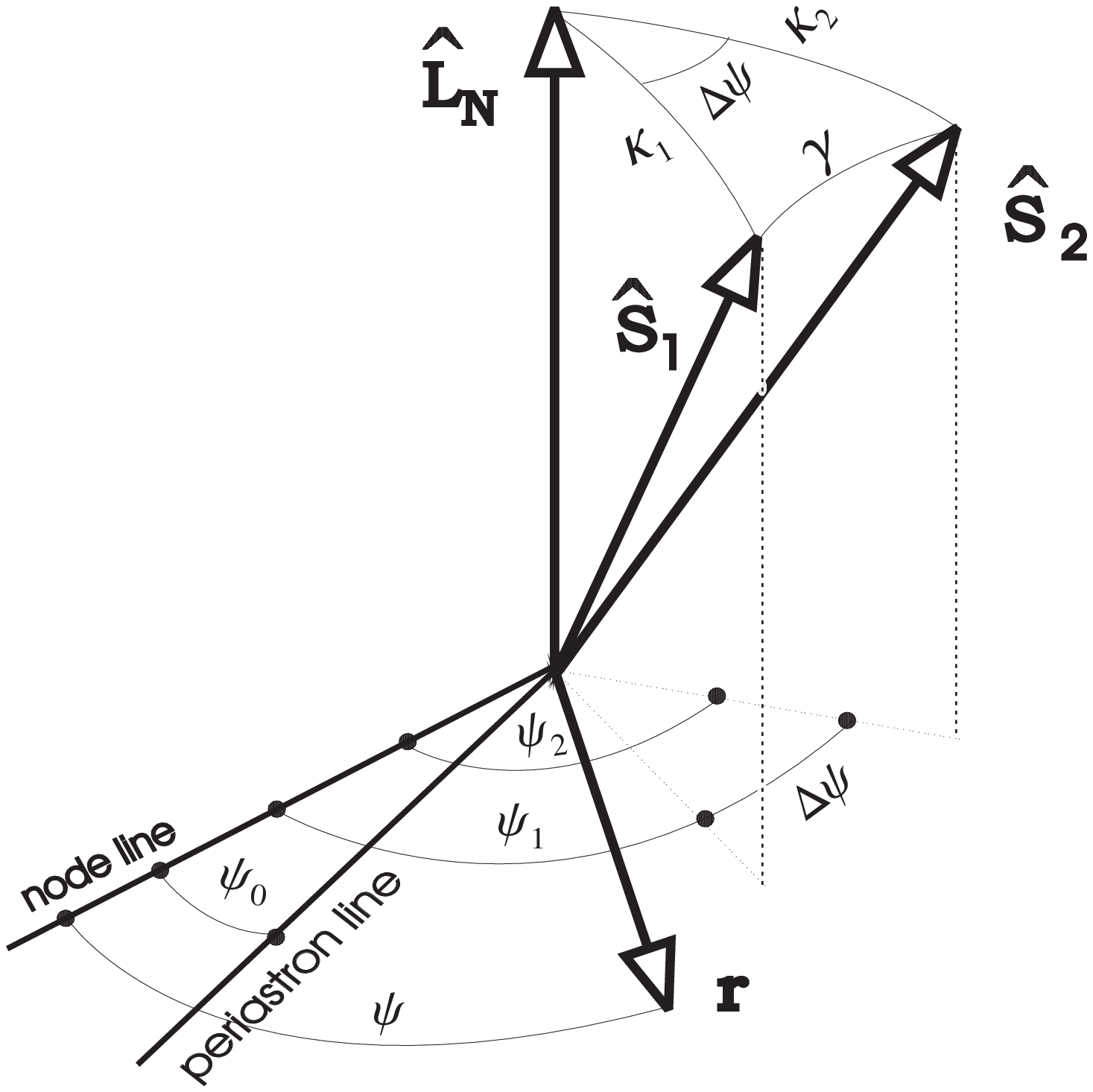}\hfill}
\caption{The angles characterizing the angular momenta, the position
{\bf r}, the direction of the periastron and the node line.}
\end{figure}

In obtaining our results we were much helped by the averaging
method on quasi-elliptical orbits developed by us. This method
relies on the application of the residue theorem for various
integrands. When written in terms of a suitably chosen parameter,
the only pole is in the origin. This feature obvously
simplifies the computations. The ,,suitably chosen
parameter" for
\vskip.25cm\hrule
\vskip3.3cm\noindent
most of the integrands is a generalization of the
true anomaly parameter $\chi$ of the Kepler orbits, defined by:
\[
\frac{2}{r}=\frac{1+\cos \chi }{r_{min}}+\frac{1-\cos \chi
}{r_{max}}\ ,
\]
where $r_{{}^{max}_{min}}$ are the values of the radial distance
$r$ at the turning points $\dot r=0$.
 
We need another parameter when computing the period. This is provided by a
generalization to the spinning binary case of the eccentric
ano\-maly parametrization of Kepler orbits. This type of
parameter was employed previously by Damour and
Deruelle \cite{DaDe} to quasi-Keplerian systems with 1PN
perturbations and by
Damour,$\!$ Sch\"afer and Wex \cite{DS1,DS2,SW,W} to the 2PN order of
accuracy. Currently we investigate under which
conditions and paramet\-rizations do the advantageous properties of the
integrands continue to remain valid \cite{param}.
 
\onecolumn
\section*{Acknowledgments}
 
This research has been supported by OTKA grants T17176 and D23744.
L.\'{A}.G was partially supported by the Hungarian State E\"{o}tv\"{o}s
Fellowship.

\end{document}